\documentclass{homework}

\usepackage{tikz} 

\usepackage[linesnumbered,ruled,vlined]{algorithm2e}
\SetKw{Continue}{continue}
\SetKw{Break}{break}

\usepackage{titlesec}

\SetCommentSty{mycommfont}

\newcommand{\bigslant}[2]{{\raisebox{.2em}{$#1$}\left/\raisebox{-.2em}{$#2$}\right.}}

\title{Finding A Path Of Length K: An Expository}
\author{Thai Bui}
\date{April 12, 2023}

\begin{document}

\maketitle

\begin{sloppypar}

\begin{abstract}
    Given a graph $G(V, E)$ and a positive integer $k$ ($k \geq 1$), a simple path on $k$ vertices is a sequence of $k$ vertices in which no vertex appears more than once and each consecutive pair of vertices in the sequence are connected by an edge. This paper provides an overview of current research on the existence and counting of k-paths in graphs.
\end{abstract}

\section{Introduction}
The $\textbf{LONGEST PATH}$ problem involves a  given graph $G$, an integer $k$ ($k \in \mathbb{N}^{\ast}$), and the task of determining if there exists a simple path of length $k$ in $G$. This problem is typically considered as a special case of the \textit{Subgraph Isomorphism} problem, which is known to be $\textbf{NP}\textit{-complete}$. Despite its computational complexity, researchers have been working on developing algorithms for finding $k$-length paths for the past thirty years. One of the brute-force approaches is to look for such a path among all subsets of $V(G)$ with cardinality $k$. This results in a $\mathcal{O}(n^k)$ algorithm, which is polynomial only when $k$ is a constant. However, Monien \cite{monien85} was the first to minimize the run-time dependency on $k$ by providing an algorithm that runs in $\mathcal{O}(k! \times \text{poly(k, n)})$. When $k \leq \frac{\log{n}}{\log{\log{n}}}$, as Williams \cite{williams05} had noted, this problem can be solved in polynomial time. It was not until Alon, Yuster, and Zwick \cite{alon95} introduced their \textit{color-coding} technique in the 1995 paper \textit{"Color-Coding"} that the case when $k$ is bounded by $\mathcal{O}\left(\log{n}\right)$ can be said to be polynomial-time solvable. They gave a randomized algorithm that runs in $\mathcal{O}\left( (2e)^k n^{\mathcal{O}(1)} \right)$ time. In 2008, Williams \cite{williams05} extended a result of Koutis \cite{koutis08} and designed an algorithm that runs in $\mathcal{O}\left( 2^k n^{\mathcal{O}(1)} \right)$ time. This paper provides an overview of the latest advancements in the field of parameterized algorithms, which have been developed based on Williams's techniques and the color-coding technique. It covers a range of versions and improvements that have been made and identifies general techniques that are commonly used. Additionally, the paper highlights some basic algorithms that demonstrate fundamental concepts in the field.

\section{Preliminaries}
\begin{enumerate}
    \item \textbf{Inclusion-exclusion principle, union version}\\
        Let $A_1, \dots , A_n$ be finite sets. Then
    $$\left| \ \bigcup_{i=1}^n \ A_i \ \right| \ = \ \sum_{\emptyset \neq X \subseteq [n]} \ (-1)^{\ |X|+1} \ . \ \left| \ \bigcap_{i \in X} \ A_i \right|$$
    \item \textbf{Inclusion-exclusion principle, intersection version}\\
    Let $A_1, \dots , A_n \ \subseteq U$, where $U$ is a finite set. Denote $\bigcap_{\ i \in \emptyset} \left( U \setminus A_i \right) = U $. Then
    $$\left| \ \bigcap_{i=1}^n \ A_i \ \right| \ = \ \sum_{X \subseteq [n]} \ (-1)^{\ |X|} \ . \ \left| \ \bigcap_{i \in X} \ \left(U \setminus A_i\right) \right|$$
\end{enumerate}

\section{A Brute-force Approach}
As mentioned in the introduction, a trivial approach to this problem would yield us an algorithm that runs in $\mathcal{O}(n^k)$ time, where $n$ is the number of vertices in $G$. This happens as a result of considering all subsets of $V(G)$ with cardinality $k$.\\ 
Monien \cite{monien85} was the first to propose a solution with slightly better performance and run in $\mathcal{O}\left(k! \ n \ |E(G)| \right)$ time. Since it is a well-known fact that a graph $G$ will contain a simple path of length $k$ if it has at least $k|V(G)|$ edges, the run time complexity of Monien's algorithm can be further simplified to $\mathcal{O}(k!\times kn^2)$. His approach sought to compute a matrix $A^{(k)}$ where the entry $a^{k}_{uv}$ is equal to \textit{some} $k$-length path from $u$ to $v$, if any such path exists, and is equal to some value $\epsilon$ otherwise. In general, instead of considering all possible paths running through a set of $k+1$ vertices, Monien only paid attention to the set of $\textit{vertices}$ lying on these $(k+1)$-paths. As it turns out, one can achieve this task by using dynamic programming and considering what is called the $\textit{q-representative}$ for the family of all paths in $G$. It is defined as follows:
\begin{definition}
(\cite{monien85}.) Let $F$ be a family of sets over $[n]$ and let $q \in \mathbb{N}$ (such that $0 \leq q < n$ ). Define $P_{j-1}(n)$ as the family of all subsets of $[n]$ of cardinality $p-1$. Then a subfamily $\Tilde{F} \subset F$ is called a \textit{q-representative} of $F$ if for all $T \in P_{\leq q}(n)$, 
$$\left(\exists U \in F \ni T \cap U = \emptyset \right) \Rightarrow \left(\exists \Tilde{U} \in \Tilde{F} \ni T \cap \Tilde{U} = \emptyset\right).$$
\end{definition}

\section{Randomized Algorithms}
We will first introduce to the readers the color coding technique (proposed by Alon, Yuster, and Zwick \cite{alon95}) and then discuss some of its improved variations (developed by Cygen et al. \cite{cygan2015parameterized} based on the work of Chen et al.\cite{chen05}), which happens when one seeks to restrict the size of the coloring. 
\subsection{Color Coding}
The color coding technique was devised by Alon, Yuster, and Zwick to provide a framework for the \textit{Subgraph Isomorphism} problem. More specifically, given two graphs $F$ and $G$ with $|V(F)| \leq |V(G)|$, the objective is to determine if there is any $X \subseteq V(G)$ such that $G[X]$ is isomorphic to $F$. We denote $\phi: V(G) \rightarrow [k]$ as a coloring of the vertices of $V(G)$.\\ 
Similar to Monien's approach of disregarding all potential paths of length $k$ passing through a set of $(k+1)$ vertices in $G$ and focusing instead on the set of vertices itself, the fundamental concept behind this technique can be defined by the following inquiry:
\begin{question}
If we are to color the vertices of $G$ uniformly at random from the color set $\set{1,\dots,k}$, what is the probability that any $X \subseteq V(G) $ with cardinality $k$ colored with pairwise distinct colors?
\end{question}
The count of colorings achievable by selecting from a set of $k$ colors for the vertices of $G$, where $|V(G)| = n$, is $k^n$. Conversely, if $X$ is a subset of $V(G)$ containing $k$ elements, then there exist $k!$ approaches to coloring $X$ such that every vertex in $X$ is assigned a unique color, and $k^{n-k}$ ways to color the remaining vertices in $V(G)\setminus X$. Hence, the probability of such an event happening is 
\begin{equation}
    \frac{k!k^{n-k}}{k^n} > \frac{1}{e^k}.
\end{equation}
We have thus reduced the problem of finding a simple path in $G$ to determining if $G$ has a colorful path of the same length, given a random coloring $\phi$. This task is in fact polynomial-time solvable as the following lemma will show:
    \begin{lemma}
    (\cite{alon95}.) Given a graph $G$ with $n$ vertices and a coloring $\phi: V(G) \rightarrow [k]$. There exists a deterministic algorithm that checks if $G$ contains a colorful path on $k$ vertices.
    \begin{proof}
    Here we introduce to the readers the more detailed form of the algorithm as provided in \cite{cygan2015parameterized}. A different version of this algorithm can be found in \textbf{\ref{appendA}}.\\
    Let $U=\set{V_1, \dots, V_k}$ be a family of sets over $V(G)$, where  $V_i = \set{u \in V(G) \ | \ \phi(u) = i}$ for any $1 \leq i \leq k$. We define $Path: S \times u \rightarrow \set{0, 1}$, where $S \subseteq [k]$ and $u \in \bigcup_{i \in S} V_i$, as follows
        \begin{equation}
            Path(S, u) = 
            \begin{cases}
                1 & \text{if}\ |S| = 1\\
                \bigvee \ \set{Path\left(\ S\setminus \{\phi(u)\},v \ \right): uv \in E(G)} & \text{if}\ \phi(u) \in S\\
                0 & \text{otherwise}
            \end{cases}
        \end{equation}
        The function $Path(S, u)$ returns $1$ if there exists a colorful path with colors from $S$ where $u$ is an endpoint and $0$ if otherwise. If such a path exists, there must be another colorful path with $v \in N(u)$ as an endpoint using all colors from $S \setminus \phi(u)$.\\
        Thus, a colorful path on $k$ vertices exists $\textit{iff}$ $Path([k], u) = 1$ for some $u \in V(G)$. Computing this function would require checking all $2^{k}$ subsets of $[k]$ and thus, the time complexity of this algorithm is $\mathcal{O}\left(2^kn^{\mathcal{O}(1)}\right)$.
    \end{proof}
    \end{lemma}
    Combining with the lower bound given for the probability of the event \textit{Question 1} concerned with, by repeating the algorithm above $e^{k}$ times, we have the following theorem:
    \begin{theorem}
        (\cite{cygan2015parameterized}.) There exists a randomized algorithm that, given a \textbf{LONGEST PATH} instance $(G, k)$, in time $(2e)^k\cdot n^{\mathcal{O}(1)}$ either reports a failure or finds a path on $k$ vertices in $G$. Moreover, if the algorithm is given a \textit{yes-instance}, it returns a solution with constant probability.
    \end{theorem}
    At this point, a natural question arises is whether changing the size of the color set would result in a better algorithm. By using $1.3k$ colors, Huffner et al. \cite{huffner08} showed that the probability in \textbf{(4.1)} can be at least $\mathcal{O}\left(1.752^{-k}\right)$. What if we reduce the size of the color set to $2$?

    \subsection{Red And Blue}
    To randomly label vertices of $G$ with only two colors, a modified approach to the LONGEST PATH problem is required. Chen et al. \cite{chen05} have suggested a new technique inspired by the Divide-and-Conquer approach. The method involves coloring each vertex of $G$ with either \textit{Red} (R) or \textit{Blue} (B) randomly and with equal probability \cite{chen05}. Let $G[R]$ be the induced subgraph in $G$ such that every vertex in this graph is colored \textit{Red}, the same thing goes for $G[B]$. Fix a simple path $P$ on $k$ vertices in $G$, if such a path exists, what we hope for with this new color assignment is that the first $\ceil{k/2}$ vertices on this path would belong to $G[R]$ and the other $\floor{k/2}$ vertices would belong to $G[B]$. It can be explicitly seen that the probability of such an event happening is $2^{-k}$ \cite{chen05}. After this step is done, we recursively apply the same procedure for $G[R]$ and $G[B]$ to reduce the size of the input instance. Here we give a summary of the framework for the Divide-and-Color technique as put together by Cygan et al. \cite{cygan2015parameterized}.
    \subsubsection{A Naive Implementation}
    For this new algorithm to work efficiently, we need to keep three things in mind:
    \begin{enumerate}
        \item The new algorithm needs to consider every possible pair of endpoints in $G$.
        \item During each recursive step, multiple random partitions must be generated and multiple recursive calls need to be made.
        \item We also need a way to store information so that the subpaths with desired properties found in $G[R]$ and $G[B]$ can be combined.
    \end{enumerate}
    The algorithm, as described in \cite{cygan2015parameterized}, proceeds as follows:\\\\
    \begin{algorithm}[H]
        \DontPrintSemicolon
        \If {$k = 1$} {
            \Return $\Tilde{M}_{S, k}[v, v] = 1$ for all $v \in S$, $0$ if $v \in V(G) \setminus S$.
        }
        With equal probability, color each vertex in $S$ with either \textit{Red} (R) or \textit{Blue} (B).\\
        $\Tilde{M}_{R, \ceil{k/2}}$ = NAIVE-COLORS-PATHS$\left(R, \ceil{k/2}\right)$ \ \CommentSty{<<R is a subset of S whose vertices are colored \textit{Red}>>}\\
        $\Tilde{M}_{B, \floor{k/2}}$ = NAIVE-COLORS-PATHS$\left(B, \floor{k/2}\right)$ \ \CommentSty{<<B is a subset of S whose vertices are colored \textit{Blue}>>}\\
        \Return $\Tilde{M}_{S, k} = \Tilde{M}_{R, \ceil{k/2}} \ \Delta \ \Tilde{M}_{B, \floor{k/2}}$
        \caption{NAIVE-COLORS-PATHS$\left(S, k\right)$}
    \end{algorithm}
    Let $S$ be a subset of $V(G)$ and $M_{S, k}(u, v)$ be a Boolean function that returns $1$ if there exists a simple path on $k$ vertices in $S$ with endpoints $u, v$. In \textit{Algorithm 1}, $\Tilde{M}_{S, k}$ is a $k \times k$ matrix that satisfies the following condition:
    \begin{equation}
        \left(\forall u, v \in S \ni u \neq v\right)\ \left(\Tilde{M}_{S, k}[u, v] = 1 \Rightarrow \ M_{S, k}(u, v) = 1\right).
    \end{equation}
    The $\Delta$ operation between two square matrices on line $6$ is defined as follows
    \begin{definition}
        (\cite{cygan2015parameterized}.) Given a colored partition $(R, B)$ of any $S \subseteq V(G)$, an $|R| \times |R|$ matrix $X$, an $|B| \times |B|$ matrix $Y$. Then $Z = X \ \Delta \ Y$ is a $|S| \times |S|$ matrix such that for any distinct pair of vertices $u, v$ of $S$:
        \begin{equation}
            Z[u, v] = 1 \Leftrightarrow u \in R \land v \in B \land \left(\exists xy \in E(G) \ni A[u ,x ] = 1 \land B[y, v] = 1\right).
        \end{equation}
    \end{definition}
    Notice that in \textbf{(4.3)}, $M_{S, k}$ is a deterministic function that reflects the existence of any simple path on $k$ vertices in $S$. The question now is what is the likelihood that the reverse \textbf{(4.3)} is true as well?\\
    To answer this question, for every subset $S$ of $V(G)$, let $\mu_{S, k}$ be the probability that, for all $u,v \in S$, if $M_{S, k}(u ,v) = 1$ then $\Tilde{M}_{S, k}[u, v] =1$. Then we define $\rho_{k}$ as follows (\cite{cygan2015parameterized})
    \begin{equation}
        \rho_{k} = \inf_{S \subseteq V(G)} \mu_{S, k}
    \end{equation}
    With probability $2^{-k}$ a simple path $P$ on $k$ vertices is correctly partitioned into two sets $R$ and $B$ (as defined above), thus we have the following inequality:
    \begin{equation}
        \rho_{k} \geq 2^{-k}\rho_{\ceil{k/2}}\ \rho_{\floor{k/2}}
    \end{equation}
    Since the depth of the recursion tree of \textit{Algorithm 1} is at most $(\log{k}-1)$ and for each $n=1,\dots, \log{k}-1$, we need to make at most $2^n$ guesses to get the correct partition of $\frac{k}{2^n}$ vertices. The following approximation of $\rho_{k}$ is given in \cite{cygan2015parameterized}:
    \begin{equation}
        \rho_{k} \approx \prod_{n=0}^{\log{k}-1} \left(2^{-k/2^n}\right) = 2^{-\mathcal{O}\left(k\log{k}\right)}
    \end{equation}
    Hence, it is necessary to repeat \textit{Algorithm 1} at least $2^{\mathcal{O}\left(k\log{k}\right)}$ times to achieve constant success probability. This would in turn make it runs significantly slower than the algorithm of \textit{Theorem 1}. So where did things start to get worse?
\subsubsection{Improved Version}
    Instead of repeating \textit{Algorithm 1} for $2^{-\mathcal{O}\left(k\log{k}\right)}$ times, we modify it to return in a single pass by repeating certain tasks several times. More specifically, the general idea of this improved version is that we want to consider more than one random partition of a subset of vertices in each pass in the hope of increasing the success probability while doing so.\\
    This new algorithm, as described in \cite{cygan2015parameterized}, is provided below:\\ \\
    \begin{algorithm}[H]
        \DontPrintSemicolon
        \If {$l = 1$} {
            \Return $\Tilde{M}_{S,\ l}[v, v] = 1$ for all $v \in S$, $0$ if $v \in V(G) \setminus S$.
        }
        Repeat the following $f(l, k)$ times
        \begin{enumerate}
            \item With equal probability, color each vertex in $S$ with either \textit{Red} (R) or \textit{Blue} (B).
            \item $\Tilde{M}_{R, \ceil{l/2}}$ = IMPROVED-COLORS-PATHS$\left(R, \ceil{l/2}\right)$ \ \CommentSty{<<R is a subset of S whose vertices are colored \textit{Red}>>}\\
            \item$ \Tilde{M}_{B, \floor{l/2}}$ = IMPROVED-COLORS-PATHS$\left(B, \floor{l/2}\right)$ \ \CommentSty{<<B is a subset of S whose vertices are colored \textit{Blue}>>}\\
            \item Compute $\Tilde{M}^{\prime}_{S,\ l} = \Tilde{M}_{R, \ceil{l/2}} \ \Delta \ \Tilde{M}_{B, \floor{l/2}}$
            \item For every $u, v \in V(G)$,\\ $\Tilde{M}_{S, \ l}[u,v] = \Tilde{M}_{S, l}[u,v] \lor \Tilde{M}^{\prime}_{S, l}[u,v]$
        \end{enumerate}
        \Return $\Tilde{M}_{S,\ l}$
        \caption{IMPROVED-COLORS-PATHS$\left(S,\ l\right)$}
    \end{algorithm}
    In what follows, we proceed to approximate the success probability of $\textit{Algorithm 2}$ by finding the appropriate computable function $f(l, k)$ on line $3$.\\
    The recursion formula for \textit{Algorithm 2} is
    \begin{equation}
        \begin{split}
            T\left(n, l, k\right) \ & \ \leq \ f(l, k)\left(T\left(n, \ceil{\frac{l}{2}}, k\right)+T\left(n, \floor{\frac{l}{2}}, k\right)\right) + n^{\mathcal{O}(1)}\\
            & \ \approx 2f(l, k)T\left(n, \frac{l}{2},k\right)+n^{\mathcal{O}(1)} 
        \end{split}
    \end{equation}
    This solves to:
    \begin{equation}
        \begin{split}
            T\left(n, l, k\right)\ & \ \approx \ 2^{\ceil{\log{l}}} \cdot \prod_{i=0}^{\ceil{\log{l}}} f\left(\frac{l}{2^i}, k\right)+\left(\sum_{i=0}^{\floor{\log{l}}}2^i\cdot \prod_{j=0}^i f\left(\frac{l}{2^j}, k\right)\right)\cdot n^{\mathcal{O}(1)}
        \end{split}
    \end{equation}
    Cygan et al. \cite{cygan2015parameterized} showed that if we choose $f(l, k)=2^{l}\log(4k)$, $\textbf{(4.10)}$ would approximate to:
    $$T(n, l, k) = 4^{l+o(l+k)}n^{\mathcal{O}(1)}$$
    A more detailed approximation for this recursion formula is given in \textbf{\ref{appendB}}.\\
    Recall that for a fixed path $P$ on $l$ vertices in $G$, the probability that $P$ is correctly partitioned into two "halves" $R$ and $B$ is exactly $2^{-l}$. Then by considering $f(l, k)$ random partitions in each pass, the probability that none of these parts contains the correct partition of $P$ is
    \begin{equation}
        \left(1-2^{-l}\right)^{f(l, k)}
    \end{equation}
    As noticed in \cite{cygan2015parameterized}, every node in the recursive tree doesn't need to represent a correct partition; it is enough for at least one of them to be accurate in each recursive call. Starting at the root node, we want at least one of the two branches represented by either $R$ or $B$ to be partitioned correctly. Since with the choice of such $f(l, k)$, the recursion tree of \textit{Algorithm 2} will have approximately $4^k$ nodes, and if we continue with this reasoning up to the leaf of the tree, there are at least $(k-1)$ nodes we have to make a right guess. Now if we substitute $f(l, k) = 2^{l}\log(4k)$ into \textbf{(4.11)}, which becomes
    $$(1-2^{-l})^{2^{l}\log(4k)} \leq \left(e^{-2l}\right)^{2^{l}\log(4k)} = \frac{1}{e^{\log(4k)}} \leq \frac{1}{2k} \leq 0.5$$
    Hence \textit{Algorithm 2} runs in $\mathcal{O}\left(4^{k+o(k)}n^{\mathcal{O}(1)}\right)$ time and if given a \textit{yes-instance}, it returns a solution with a constant probability \cite{cygan2015parameterized}.

\section{Algebraic Techniques}
This section introduces two algebraic techniques that result in 
two randomized $\textbf{LONGEST PATH}$ algorithms that run in $\mathcal{O}\left(4.32^k \cdot k \cdot n^2\right)$ (\cite{amini11}) and $\mathcal{O}\left(2^k \cdot n^{\mathcal{O}(1)}\right)$ (\cite{williams05}) time, respectively.
\subsection{Prerequisites}
    Given two graphs $P$ and $G$, a graph $\textit{homomorphism}$ from $P$ to $G$ is a function $f$ from $V(P)$ to $V(G)$ such that
    $$uv \in E(P) \Rightarrow f(u)f(v) \in E(G).$$
    When $f$ is injective, $f$ is called an $\textit{injective homomorphism}$. Using the notations introduced by Amin et al. \cite{amini11}, we also have the following four definitions:
    \begin{enumerate}
        \item Hom($P, G$): the number of homomorphisms from $P$ to $G$.
        \item Inj($P, G$): the number of injective homomorphisms from $P$ to $G$.
        \item Sub($P, G$): the number of distinct copies of $P$ contained in $G$.
        \item Aut($P$): the number of injective homomorphisms from $P$ to itself. In other words, the number of ways one can map an object into itself while preserving its structure. 
    \end{enumerate}
\subsection{Counting Paths}
    The first algorithm, proposed by Amini et al. \cite{amini11}, seeks to count the number of simple paths on $k$ vertices in $G$. The crux of this method depends significantly on the following algorithm 
    \begin{theorem}
        (\cite{diaz00}.) Let $P$ be a path graph on $k$ vertices and $G$ be a graph on $n$ vertices respectively. Then Hom($P, G$) is computable in time $\mathcal{O}\left(k \cdot n^2\right)$ and space $\mathcal{O}\left(\log{k} \cdot n^2\right)$.
    \end{theorem}
    Let $P_k$ be the path graph on $k$ vertices. To compute the number of distinct copies of $P_k$ in $G$, we have the following simple formula \cite{amini11}
    \begin{equation}
        \text{Sub}(P_{k}, G) = \frac{\text{Inj}(P_{k}, G)}{\text{Aut}(P)} = \frac{\text{Inj}(P_{k}, G)}{2}
    \end{equation}
    Since Aut($P_k$) = $2$ for every simple path on $k$ vertices, it suffices to calculate the number of \textit{injective homomorphisms} from $P_k$ to $G$ \cite{amini11}.\\ 
    \begin{question}
        Let $F$ and $G$ be two graphs on $k$ and $n$ vertices respectively, how can we calculate $\text{Inj}(F, G)$?
    \end{question}
    It turns out that with the help of the \textit{Inclusion-Exclusion principle}, we can effectively deduce a formula for Inf($F, G$). First, let's limit our formula for the case when $k=n$. 
    \begin{lemma}
        (\cite{amini11}.) Let $F$ and $G$ be two graphs with $|V(F)| = |V(G)|$. Then
        \begin{equation}
            \text{Inj}(F, G) = \sum_{W \subseteq V(G)} (-1)^{\ |W|}\ \text{Hom}(F, G\setminus W)
        \end{equation}
        \begin{proof}
            Here we show a different proof for this lemma other than the one provided in \cite{amini11}.\\
            Let the universe $U$ be the set of all \textit{homomorphisms} from $F$ to $G$. For every vertex $u \in V(G)$, let $A_u$ be the set of homomorphisms from $F$ to $G$ such that 
            $$\left(\forall f \in A_u\right)\left(|f^{-1}(u)| \geq 1\right)$$
            Thus $A_u$ is the set of homomorphisms from $F$ to $G$ that have $u$ in their images. This means that the union $\bigcap_{u \in V(G)} A_u$ is the set of all injective homomorphisms from $F$ to $G$.\\
            Applying the Inclusion-Exclusion principle,
            $$|\bigcap_{u \in V(G)} A_u| = \sum_{W \subseteq V(G)}\ (-1)^{\ |W|}|\bigcap_{u \in W} U \setminus A_u|.$$
            By the definition of $A_u$, for every $W \subseteq V(G)$, $\bigcap_{u \in W} U \setminus A_u$ is the set of homomorphisms from $F$ to $G$ that doesn't have $W$ in their images. Hence we have
            $$\text{Inj}(F, G) = \sum_{W \subseteq V(G)} (-1)^{\ |W|}\ \text{Hom}(F, G\setminus W)$$
        \end{proof} 
    \end{lemma}
    For every $W \subseteq V(G)$, let $W^{\prime} = G\setminus W$, then \textbf{(5.2)} can be written as follows (\cite{amini11}):
    $$\text{Inj}(F, G) = \sum_{W^{\prime} \subseteq V(G)} (-1)^{\ |V(G)|- |W^{\prime}|}\ \text{Hom}(F, G[W^{\prime}])$$
    It is not hard to see that in the case where $n \geq k$, to compute the number of injective homomorphisms from $F$ to $G$, we have to calculate the total number of injective homomorphisms from $F$ to every induced subgraph $G^{\prime}$ on $k$ vertices of $G$. The following theorem provides a compact form of such a formula:
    \begin{theorem}
        (\cite{amini11}.) Let $F$ and $G$ be two graphs on $k$ and $n$ vertices respectively, where $k \leq n$. Then
        $$\text{Inj}(F, G) = \sum_{Y \subseteq V(G), |Y| \leq k} (-1)^{k-|Y|} \binom{n-|Y|}{k-|Y|} \text{Hom}(F, G[Y])$$
    \end{theorem}
    Combining $\textit{Theorem 3}$ with the algorithm in $\textit{Theorem 2}$, a deterministic algorithm that counts the number of simple paths on $k$ vertices in $G$ will run in time 
    $$\mathcal{O}\left(k \cdot \sum_{i = 0}^{k}i^2 \cdot \binom{n}{i}\right)$$
    and space $\mathcal{O}(\log{k}\cdot n^2)$.\\
    A further investigation into the time complexity of the algorithm above yields us
    \begin{equation}
        \begin{split}
           \mathcal{O}\left( k \cdot \sum_{i = 0}^{k}i^2 \cdot \binom{n}{i}\right) & \ = \ \mathcal{O}\left(k \cdot \sum_{i = 0}^{n}i^2 \cdot \binom{n}{i}\right)\\
            & \ = \ \mathcal{O}\left(2^{n-2}n(n+1)k\right)\\
            & \ = \ \mathcal{O}\left(2^{n-2} \cdot k\cdot n^2\right)
        \end{split}
    \end{equation}
    As we can see, this deterministic algorithm runs in exponential time in $n$. This is due to the fact that to obtain the total number of injective homomorphisms from $F$ to $G$, we need to count the number of injections from $F$ to every subset of $k$ vertices in $V(G)$. This in turns creates a lot of overhead in our computation if we just want to know if there is a copy of $F$ in $G$.\\
    A randomized version of this algorithm unburdens the excessive workload by restricting the cardinality of the family of subsets that we will take into consideration and will again utilize the color-coding technique introduced in section $\textit{4.1}$ (\cite{amini11}). Given two graphs $F$ and $G$ on $k$ and $n$ vertices respectively, a coloring $\phi: V(G) \rightarrow [k^{\ast}]$, we denote \textit{col-Inj($F, G$)} as the number of colorful injective homomorphisms of $F$ in $G$. Let $G^{\#}$ be the graph obtained from $G$ by deleting edges with endpoints of the same color. Amini et  al. \cite{amini11} gave the following observation:
    $$\text{col-Inj}(F, G) = \text{col-Inj}(F, G^{\#})$$
    If we use $k^{\ast} = 1.3k$ colors to label $V(G)$, the probability that a simple path on $k$ vertices in $G$ becomes colorful is at least $1.752^{-k}$ (\cite{huffner08}). Thus, using this new set of colors, we can repeat this new randomized algorithm $1.752^{k}$ times to obtain constant success probability. The question now is how should we develop our algorithm to reduce the overhead of the deterministic one in each pass.\\
    Using the same notations as in \textit{Lemma 1}, we denote $V^{\#}_i = \set{  u \in V(G^{\#}) \ | \ \phi(u) = i }$. Considering the family $U = \{V^{\#}_1, \dots V^{\#}_{k^{\ast}}\}$, each element of this family is an independent set in $G^{\#}$ since any edges whose endpoints are of the same color have been removed (\cite{amini11}). With this new insight, given a coloring $\phi: V(G) \rightarrow [1.3k]$, instead of considering every subset of $k$ vertices in $V(G^{\#})$, we will repeatedly run our formula through $2^{1.3k}$ subsets of the family $U$ (\cite{amini11}). The new formula in this scenario is as follows
    \begin{theorem}
        \text{(\cite{amini11}.)} Let $\phi: V(G) \rightarrow [k^{\ast}]$ be a coloring of $G$, $k^{\ast} \geq k$, and $V^{\#}_i = \phi^{-1}(i)$. Then
        $$\text{col-Inj}(F, G) = \text{col-Inj}(F, G^{\#}) = \sum_{I \subseteq [k^{\ast}]; |I| \leq k} (-1)^{k-|I|} \binom{k^{\ast}-|I|}{k-|I|} \text{Hom}\left(F, G^{\#}\left[\bigcup_{i \in I} V^{\#}_i\right]\right)$$
    \end{theorem}
    At each pass, by letting $k^{\ast} = 1.3k$, our algorithm runs in time 
    $$\mathcal{O}\left(k\cdot \sum_{i=0}^{k} i^2 \cdot \binom{1.3k}{i}\right) = \mathcal{O}\left(2^{1.3k}\cdot k \cdot n^2\right)$$
    Hence if we repeat this process $1.752^k$ times, the final randomized algorithm will run in time
    $$\mathcal{O}\left(4.32^k \cdot k \cdot n^2\right)$$
    and space $\mathcal{O}(\log{k}\cdot n^2)$. This algorithm will return \textit{YES} if at any point during all these $1.752^k$ passes we have col-Inj($F, G$) $> 0$ \cite{amini11}.

\subsection{Counting Over Field Of Characteristic 2}
    In the previous section, we introduced a formula for counting the number of simple paths on $k$ vertices in a graph $G$ over $(\mathbb{N}, +, \times)$ (the semiring of natural numbers). In what follows, we will proceed to construct an algebraic structure for \textit{counting} the same object of our inquiry over a finite field of characteristic $2$. The first $\mathcal{O}\left(2^{k}\cdot n^{\mathcal{O}(1)}\right)$-time algorithm was an improvement made by Williams \cite{williams05} based on Koutis's $\mathcal{O}\left(2^{1.5k}\cdot n^{\mathcal{O}(1)}\right)$-time algorithm \cite{koutis08}. Here we introduce to the reader the framework provided by Cygan et al. \cite{cygan2015parameterized} in the case where $G$ is a directed graph. Let us tackle this new technique bit by bit.
    \subsubsection{Main Tools}
    The first tool that we are going to utilize in this section is a \text{finite field} $(F, +, \times)$. A field is a set on which addition and multiplication are defined and behave as the corresponding operations on rational and real numbers. This finite field $F$ is a field that contains a finite number of elements. Moreover, the following axioms are needed to be satisfied:
    \begin{enumerate}
        \item  \textbf{Closure}: For any two elements $a,b$ in the field, $a+b$ and $ab$ must also be     elements of the field.
        \item \textbf{Associativity}: For any three elements a,b,c in the field, $(a+b)+c = a+(b+c)$ and $(ab)c = a(bc)$.
        \item \textbf{Inverse elements}: For any \textit{non-zero} element a in the field, there exists an element $-a$ such that $a+(-a) = 0$ and an element $a^{-1}$ such that $aa^{-1} = 1$.
        \item \textbf{Distributivity}: For any three elements a,b,c in the field, $a(b+c) = ab+ac$ and $(a+b)c = ac+bc$.
    \end{enumerate}
    Specifically, our field of interest is a finite field of characteristic $2$ with order $s$ (where $s$ is an integer); we denote this field as $\mathbb{F}_{2}$. Since $2$ is a prime number, it is a well-known fact that $\mathbb{F}_2$ is unique up to isomorphism.\\
    Given a field $F$ and any vector $\Vec{x} \in F^n$ where each component of $\Vec{x} = \left(\Tilde{x}_1, \dots, \Tilde{x}_n\right)$ is an element of $F$, a multivariate polynomial $p(\Vec{x})$ is a mathematical expression that contains a finite number of monomial terms, each monomial term has the form of $a_{c_1, \dots, c_n}\prod_{i=1}^n \Tilde{x}^{c_i}_i$ where each $c_i$ is in $\mathbb{N}^{\ast}$ and $a_{c_1, \dots, c_n} \in F$ is called a coefficient.
    $$p(\Vec{x}) = \sum_{\{c_i\}^{ n}_{i=1} \in \ (\mathbb{N} \cup \ 0)^n} a_{c_1, \dots, c_n}\prod_{i=1}^n \Tilde{x}^{c_i}_i$$
    Every polynomial $p$ defined above is in turn an element of the ring $F[x_1, \dots, x_n]$.\\
    The second tool that we also need to use is the \textit{Schwartz–Zippel lemma}, which states
    \begin{lemma}
        Let $P \in F[x_1, \dots, x_n]$ be a \textit{non-zero} polynomial of total degree $d \geq 0$ over a field $F$. Let $S$ be a finite subset of $F$ and let $r_1,\dots,r_n$ be selected at random independently and uniformly from $S$. Then
        $$\text{Pr}\left[P\left(r_1,\dots,r_n\right)=0\right] \leq \frac{d}{|S|}$$
    \end{lemma}
\subsubsection{New Representation For Paths}
    Given a directed graph $G$ and an integer $k$, a $\textit{k-walk}$ in $G$ is either an overlapping path or a simple path on $k$ vertices. The main objective of this method is to construct a polynomial that characterizes every possible k-walk in $G$ and $G$ only contains a k-path if and only if this polynomial is non-zero.\\
    Every k-path in $G$ is a sequence of vertices and edges and can be seen as follows:
    $$v_1,e_1, \dots ,e_{k-1},v_k$$
    If a k-walk is overlapped, it can also be seen as a \textit{non-injective} homomorphism from $P_k$ to $G$. Otherwise, it is an injective one. If one is to characterize every \textit{k-path} with a monomial of the form
    $$\prod_{i=1}^{k-1}x_{e_i} \cdot \prod_{i=1}^{k}y_{v_i}$$
    , according to Cygan et al. \cite{cygan2015parameterized}, the expression not only encompasses all simple paths on $k$ vertices in $G$ but also takes into account all the undesirable overlapping k-walks. To separate one category from the other, one can improve the descriptiveness of such a sequence of k-walks by labeling them using a bijection $l: [k] \rightarrow [k]$. More specifically, for every k-walk $W=v_1,\dots, v_k$ in $G$ and a bijection $l$, we define the monomial
    \begin{equation}
        \prod_{i=1}^{k-1} x_{v_i,\ v_{i+1}} \cdot \prod_{i=1}^{k} y_{v_i,\ l(i)}
    \end{equation}
    By following this approach, each vertex $v_i$ is assigned a unique label. In case our walk visits a vertex $v_i$ multiple times, every time it overlaps with the vertex, $v_i$ is given a distinct label.\\ 
    With this definition for each k-walk, we are ready to define our polynomial $P$ (\cite{cygan2015parameterized})
    \begin{equation}
        P(\Vec{\textbf{x}}, \Vec{\textbf{y}}) = \sum_{\text{walk $W=v_1,\dots,v_k$}}\ \sum_{\substack{l: [k] \rightarrow [k]\\\text{l is bijective}}} \prod_{i=1}^{k-1} x_{v_i,\ v_{i+1}} \cdot \prod_{i=1}^{k} y_{v_i,\ l(i)}
    \end{equation}
    where $\Vec{\textbf{x}}$ and $\Vec{\textbf{y}}$ are vectors of all variables in $x$ and $y$, respectively. With this setup ready, the remaining task is to decide if $P$ is identically zero over the finite field $\mathbb{F}_2$ of size $2^{\ceil{\log{(4k)}}}$ \cite{cygan2015parameterized}.\\
    In \textbf{5.2}, we want to know if Inj($F, G$) (or col-Inj($F, G$)) is greater than $0$. We accomplished this by subtracting the number of \text{non-injective} homomorphisms from $F$ to $G$ from Hom($F, G[Y]$) for every $Y \subseteq V(G)$ with cardinality $k$. But to compute such quantities we have to calculate Hom$(F, G[Y^{\prime}])$ of every subset $Y^{\prime}$ of $V(G)$ whose size is less than $k$. Such a tedious task is conveniently achieved as we construct the multivariate polynomial $P\left(\Vec{\textbf{x}}, \Vec{\textbf{y}}\right)$ since every non-path walk $W$ is canceled over $\mathbb{F}_{2}$.
    \begin{lemma}
        \text{(\cite{cygan2015parameterized}.)}\ \ $P(\Vec{\textbf{x}}, \Vec{\textbf{y}}) = \sum_{\text{path $W=v_1,\dots,v_k$}}\ \sum_{\substack{l: [k] \rightarrow [k]\\\text{l is bijective}}} \prod_{i=1}^{k-1} x_{v_i,\ v_{i+1}} \cdot \prod_{i=1}^{k} y_{v_i,\ l(i)}$
    \end{lemma}
    The proof for this canceling property of $P$ is based on the observation that given a k-walk $W$ and a bijection $l$, one can always pair it up with another bijection $l^{\prime}$ such that two monomials associated with $l$ and $l^{\prime}$ are the same. A different proof for \textit{Lemma 4} is provided in \textbf{\ref{appendC}} (the main idea here is to partite the set of all monomials associated with a walk $W$ into appropriate equivalence classes and then to show that each equivalence class sums up to $0$ over $\mathbb{F}_2$).\\ 
    As a result $P(\Vec{\textbf{x}}, \Vec{\textbf{y}})$ is not identically zero if and only if $G$ contains a k-path.\\
    By using the weighted version of the Inclusion-Exclusion principle, Cygan et al. \cite{cygan2015parameterized} derived the following formula for calculating $P(\Vec{\textbf{x}}, \Vec{\textbf{y}})$ at a given pair $(\Vec{\textbf{x}}, \Vec{\textbf{y}})$
    \begin{equation}
        P(\Vec{\textbf{x}}, \Vec{\textbf{y}}) = \sum_{X \subseteq [k]}\ \sum_{\text{walk $W$}}\ \sum_{\substack{l: [k] \rightarrow X\\\text{l is bijective}}} \prod_{i=1}^{k-1} x_{v_i,\ v_{i+1}} \cdot \prod_{i=1}^{k} y_{v_i,\ l(i)}
    \end{equation}
    Given $X \subseteq [k]$, denote $P_{X}(\Vec{\textbf{x}}, \Vec{\textbf{y}}) = \sum_{\text{walk $W$}}\ \sum_{\substack{l: [k] \rightarrow X\\\text{l is bijective}}} \prod_{i=1}^{k-1} x_{v_i,\ v_{i+1}} \cdot \prod_{i=1}^{k} y_{v_i,\ l(i)}$ (\cite{cygan2015parameterized}). Then by using dynamic programming, 
    \begin{lemma}
        (\cite{cygan2015parameterized}.) Let $X \subseteq [k]$. The polynomial $P_X$ can be evaluated using $\mathcal{O}(km)$ field operations.
    \end{lemma}
    Next, the algorithm carries out the following steps \cite{cygan2015parameterized}:
    \begin{enumerate}
        \item Pick a vector of random elements from the field $\mathbb{F}_2$ with size $2^{\ceil{\log{(4k)}}} \geq 4k$.
        \item Evaluate the polynomial $P$ over this vector.
        \item Return NO if we get $0$. Since the degree of $P$ is at most $2k-1$, applying the Schwartz-Zippel lemma we have that the probability that $P$ is identically zero is at most $\frac{1}{2}$.
        \item Otherwise, return YES.
    \end{enumerate}
    Overall, this algorithm runs in $\mathcal{O}\left(2^{k}\cdot n^{\mathcal{O}(1)}\right)$ time. Moreover, as pointed out by Cygan et al. \cite{cygan2015parameterized}, applying the technique outlined in Lemma 5 enables us to enhance the algorithm described in Lemma 1, allowing it to operate within polynomial space.

\section{Conclusion}
In conclusion, the \textbf{LONGEST PATH} problem is a challenging task of determining if a simple path of length $k$ exists in a given graph $G$. It should be noted that, while all of the randomized algorithms presented, which are based on some variant of the color coding technique, can be efficiently derandomized using a splitter pseudorandom object, there is currently no known deterministic version of the algorithm described in the section \textbf{5.3}. This algorithm is classified among those related to the polynomial identity testing problem, and the absence of a deterministic variant remains a major unresolved issue in the field.

\appendix
\section{Improving The Space Complexity For Lemma 1} \label{appendixA}
\label{appendA}
We show that the algorithm in \textit{Lemma 1} can be optimized to execute in polynomial space.\\ 
Given a graph $G$ on $n$ vertices, an integer $k$, and a coloring $\phi: V(G) \rightarrow [k]$, it can be explicitly seen that any k-walk whose vertices are colored with all colors from $[k]$ is a colorful k-path.\\
For all $i \in [k]$, we define $A_i$ as the set of all k-walks in $G$ that pass through at least a vertex of the $i^{\text{th}}$ color. This means that $\bigcap_{i \in [k]} A_i$ is the set of all colorful k-paths in $G$. Let $U$ be the set of all k-walks in $G$, then by applying the Inclusion-Exclusion principle,
$$|\bigcap_{i \in [k]} A_i| = \sum_{X \subseteq [k]} (-1)^{|X|} |\bigcap_{i \in X} (U \setminus A_i)|$$
For every $X \subseteq [k]$, the right-hand side of the above equation suggests that we can determine the existence of a k-path in $G$ just by counting the number of k-walks that have vertices colored with colors of $[k] \setminus X$ only.\\
Let $S \subseteq [k]$ be arbitrary, using dynamic programming, we fill in a two-dimensional array $C$ where $C[v, j]$ denotes the number of j-walks that start at $v$ using colors from $S^{\prime} = [k] \setminus S$ only. Let $\alpha$ be a function of two variables such that, for every pair of vertices $u, w \in V(G)$, $\alpha(v, w)$ returns $1$ if $vw \in E(G)$ and $0$ otherwise. Then:
\begin{equation}
    C[v, j] = 
    \begin{cases}
        \sum_{h \in S^{\prime}} \sum_{w \in \phi^{-1}(h)}\alpha(v, w) & \text{if}\ j = 1\\
        \sum_{h \in S^{\prime}} \sum_{w \in \phi^{-1}(h)}\alpha(v, w)T[w, j-1] & \text{otherwise}
    \end{cases}
\end{equation}
In the above recurrence, the number of k-walks that have vertices colored with colors of $S^{\prime}$ only is $\sum_{v \in V(G)} C[v, k]$. Thus, determining $|\bigcap_{i \in S} (U \setminus A_i)|$ can be done in $\mathcal{O}\left(k \cdot n^2 \right)$ time.\\
Hence calculating $|\bigcap_{i \in [k]} A_i|$ can be done in $\mathcal{O}\left(2^k \cdot \textit{poly}(n)\right)$ time and polynomial space.

\section{Approximation For 4.10}
\label{appendB}
Here we show that \textbf{(4.10)} is approximated to $4^{l+o(l+k)}n^{\mathcal{O}(1)}$. 
Recall that
$$T(n, l, k) \approx 2^{\ceil{\log{l}}}\cdot \prod_{i=0}^{\ceil{\log{l}}} f\left(\frac{l}{2^i}, k\right)+\left(\sum_{i=0}^{\floor{\log{l}}}2^i\cdot \prod_{j=0}^i f\left(\frac{l}{2^i}, k\right)\right)\cdot n^{\mathcal{O}(1)}$$
If we choose $f(l, k) = 2^l \log{(4k)}$, then
\begin{equation}
    \begin{split}
        T(n, l, k) & \ \approx \  2^{\ceil{\log{l}}}\cdot \prod_{i=0}^{\ceil{\log{l}}} 2^{l/2^i} \log{(4k)}+\left(\sum_{i=0}^{\floor{\log{l}}}2^i\cdot \prod_{j=0}^i 2^{l/2^j}\log{(4k)}\right)\cdot n^{\mathcal{O}(1)}\\
    \end{split}
\end{equation}
Let $K = 2^{\ceil{\log{l}}}\cdot \prod_{i=0}^{\ceil{\log{l}}} 2^{l/2^i} \log{(4k)}$, then
\begin{equation}
    \begin{split}
        K & \ = \  2^{\ceil{\log{l}}}\cdot 2^l \cdot 2^{\sum_{i=0}^{\ceil{\log{l}}}\frac{1}{2^i}} \cdot (\log{(4k)})^{\ceil{\log{l}}+1}\\
        & \ \leq \ 2^{\ceil{\log{l}}}\cdot 2^l \cdot 2^2 \cdot \left(\log{(4k)}\right)^{\ceil{\log{l}}} \cdot \log{(4k)}\\
        & \ = \ 4^l \cdot \left(2\log{(4k)}\right)^{\ceil{\log{l}}} \cdot 4^{\mathcal{O}\left(\log{\log{k}}\right)}\\
        & \ = \ 4^l \cdot 4^{\mathcal{O}\left(\log{\log{k}}\right)\cdot \ceil{\log{l}}} \cdot 4^{\mathcal{O}\left(\log{\log{k}}\right)}\\
        & \ = \ 4^{l+\mathcal{O}\left(\log{\log{k}}\right)(1+\ceil{\log{l}})} = 4^{l+o(l+k)}
    \end{split}
\end{equation}
Similarly, 
\begin{equation}
    \begin{split}
        \left(\sum_{i=0}^{\floor{\log{l}}}2^i\cdot \prod_{j=0}^i 2^{l/2^j}\log{(4k)}\right)\cdot n^{\mathcal{O}(1)} & \ \leq \ \left(\sum_{i=0}^{\floor{\log{l}}}2^i\cdot \prod_{j=0}^{\floor{\log{l}}} 2^{l/2^j}\log{(4k)}\right)\cdot n^{\mathcal{O}(1)}\\
        & \ \leq \ \left(\prod_{j=0}^{\floor{\log{l}}} 2^{l/2^j}\log{(4k)}\right) \cdot \sum_{i=0}^{\floor{\log{l}}} 2^i \cdot n^{\mathcal{O}(1)}\\
        & \ \leq \ \left(\prod_{j=0}^{\floor{\log{l}}} 2^{l/2^j}\log{(4k)}\right) \cdot 2^{\ceil{\log{k}}} \cdot n^{\mathcal{O}(1)}\\
        & \ \leq \ K \cdot n^{\mathcal{O}(1)}\\
        & \ \leq \ 4^{l+o(l+k)} \cdot n^{\mathcal{O}(1)}
    \end{split} 
\end{equation}
Combining \textbf{(6.2)} with \textbf{(6.3)} completes our approximation. 

\section{A Different Proof For Lemma 4}
\label{appendC}
We give a slightly different proof for \textit{Lemma 4} other than the one provided in \cite{cygan2015parameterized}.
\begin{proof}
Let $W=v_1, \dots, v_k$ be an arbitrary k-walk in $G$. Based on the equation of \textbf{(5.5)}, the sub-polynomial that represents $W$ is as follows:
\begin{equation}
    \begin{split}
        \Tilde{w}\left(\Vec{x}, \Vec{y}\right) \ & \ = \ \sum_{\substack{l: [k] \rightarrow [k]\\\text{l is bijective}}} \prod_{i=1}^{k-1} x_{v_i,\ v_{i+1}} \cdot \prod_{i=1}^{k} y_{v_i,\ l(i)}\\
        & \ = \ \prod_{i=1}^{k-1} x_{v_i,\ v_{i+1}} \cdot \left(\sum_{\substack{l: [k] \rightarrow [k]\\\text{l is bijective}}} \prod_{i=1}^{k} y_{v_i,\ l(i)}\right)\\
        & \ = \ \prod_{i=1}^{k-1} x_{v_i,\ v_{i+1}} \cdot \left(\sum_{\sigma \in S_k} \prod_{i=1}^{k} y_{v_i,\ \sigma(i)}\right)\\
    \end{split}
\end{equation}
where $S_k$ is the permutation group of the set $[k]$.\\
Let $W^{\prime}$ be the $k$-tuple whose elements are vertices on $W$, i.e. $W^{\prime} = (v_1, \dots, v_k )$, and we denote $Perm(W^{\prime})$ as the set of all permutations of $W^{\prime}$ under the action of $S_k$. It can be explicitly seen that each monomial $\prod_{i=1}^{k} y_{v_i,\ \sigma(i)}$ (where $\sigma \in S_k$) can be associated with one and only one element of $Perm(W^{\prime})$. Since $W$ is a k-walk, there is at least one vertex $v_i$ ($1 \leq i \leq k$) that appears many times on $W$. Let $U=\{j_1, \dots ,j_r\}$ be the set of indices such that $\forall w \in U, v_w = v_i $. Clearly, $i \in U$ and $|U| \geq 2$. For each $\sigma \in S_k$, let $U^{\sigma} = \{\sigma(j_1), \dots, \sigma(j_r)\}$. For each $a \in Perm(W^{\prime})$, let $\sigma_{a}$ be the element of $S_k$ such that $\sigma_{a}\left(W^{\prime}\right) = a$. We define the relation $\simeq$ on $Perm(W^{\prime})$ as follows:
$$\left(\forall a,b \in Perm(W^{\prime})\right)\left( a \simeq b \Leftrightarrow U^{\sigma_a} = U^{\sigma_b}\right)$$
Thus $\simeq$ is an equivalence relation and for each equivalence class $\mathcal{C}$ of $\bigslant{Perm\left(W^{\prime}\right)}{\simeq}$, we will prove that $\sum_{c \in \mathcal{C}} \prod_{i=1}^k y_{v_i, \sigma_c(i)}$ is equal to $0$ over the field $\mathbb{F}_2$.\\
To see this, let $\mathcal{C} \in \bigslant{Perm\left(W^{\prime}\right)}{\simeq}$ be arbitrary, we then again define the relation $\leftrightarrow$ on $\mathcal{C}$ as follows
$$\left(\forall a, b \in \mathcal{C}\right)\left(a \leftrightarrow b \Leftrightarrow \left(\forall h \in [k]\setminus U, \sigma_{a}(h) = \sigma_{b}(h)\right)\right)$$
It is not hard to see that $\leftrightarrow$ is an equivalence relation. Let $\mathcal{Z} \in \bigslant{\mathcal{C}}{\leftrightarrow}$ be arbitrary then 
\begin{equation}
    \begin{split}
        \sum_{z \in \mathcal{Z}} \prod_{i=1}^k y_{v_i, \sigma_{z}(i)} & \ = \ \sum_{z \in \mathcal{Z}} \prod_{i \in [k]\setminus U} y_{v_i, \sigma_{z}(i)} \cdot \prod_{i \in U} y_{v_i, \sigma_{z}(i)}\\
        & \ = \ \prod_{\substack{i \in [k] \setminus U\\ \text{any $z \in \mathcal{Z}$}\\ }} y_{v_{i}, \sigma_{z}(i)} \cdot \left(\sum_{z \in \mathcal{Z}}\prod_{i \in U} y_{v_i, \sigma_{z}(i)}\right) 
    \end{split}
\end{equation}
There are exactly $|U|!$ terms in $\sum_{z \in \mathcal{Z}}\prod_{i \in U} y_{v_i, \sigma_{z}(i)}$ and for any two $\Tilde{z}, \Hat{z} \in \mathcal{Z}$, the two monomials $\prod_{i \in U} y_{v_i, \sigma_{\Tilde{z}}(i)}$ and $\prod_{i \in U} y_{v_i, \sigma_{\Hat{z}}(i)}$ are the same. To see this, for each $m \in U$, there is a unique $n \in U$ such that $\sigma_{\Tilde{z}} (m) = \sigma_{\Hat{z}}(n)$ and this also means $y_{v_m, \sigma_{\Tilde{z}}(m)} = y_{v_n, \sigma_{\Hat{z}}(n)}$. Since there are an even number of terms ($|U| \geq 2$) in $\sum_{z \in \mathcal{Z}}\prod_{i \in U} y_{v_i, \sigma_{z}(i)}$, the expression $\sum_{z \in \mathcal{Z}} \prod_{i=1}^k y_{v_i, \sigma_{z}(i)}$ is equal to $0$ over $\mathbb{F}_2$. Thus $\sum_{c \in \mathcal{C}} \prod_{i=1}^k y_{v_i, \sigma_c(i)}$ is equal to $0$ over the field $\mathbb{F}_2$.\\ 
This completes our proof.
\end{proof}

\end{sloppypar}
\end{document}